\documentclass[aps,epsf,epsfig,twocolumn]{revtex4}
\usepackage[dvips]{graphicx}
\usepackage[usenames]{color}
\usepackage{amssymb}
\usepackage{amsmath}

\def\be{\begin{equation}}
\def\ee{\end{equation}}
\def\bea{\begin{eqnarray}}
\def\eea{\end{eqnarray}}
\def\bi{\begin{itemize}}
\def\ei{\end{itemize}}
\def\bin{\begin{enumerate}}
\def\ein{\end{enumerate}}
\def\la{\langle}
\def\ra{\rangle}

\newcommand{\K}{{\bf k}}
\newcommand{\vect}[1]{\mathbf{#1}}

\begin{document}

\title{Second order quantum phase transition of a homogeneous Bose gas with attractive interactions}

\author{P. Zi\'n,$^1$ B. Ole\'s,$^2$
M. Trippenbach,$^{1,3}$ and K. Sacha$^2$}

\affiliation{$^1$Soltan Institute for Nuclear Studies, Ho\.za 69,
00-681 Warsaw, Poland}

\affiliation{$^3$Marian Smoluchowski Institute of Physics and
Mark Kac Complex Systems Research Centre, Jagiellonian University,
Reymonta 4, 30-059 Krak\'ow, Poland}

\affiliation{$^2$Institute for Theoretical Physics, Warsaw University,
Ho\.za 69, 00-681 Warsaw, Poland}

\date{\today}

\begin{abstract}
We consider a homogeneous Bose gas of particles with an attractive interaction. Mean field theory predicts for this system a spontaneous symmetry breaking at a certain value of the interaction strength. We show that at this point a second-order quantum phase transition occurs. We investigate the system in the vicinity of the critical point using Bogoliubov theory and a continuous description, that allows us to analyze {\it quantum fluctuations} in the system even when the Bogoliubov approach breaks down.
\end{abstract}

\maketitle
\section{Introduction}

Many body problems in a quantum degenerate limit can be investigated experimentally and theoretically in ultra cold atomic gasses \cite{Leggett,castin,stringari}. At the current level of the experimental control and detection accuracy various phenomena and effects predicted theoretically (for example Mott-Insulator transition \cite{bloch}) can now be precisely investigated and tested. The most fascinating examples from our point of view consists of BEC-BCS transition \cite{BCS} and more recently critical behavior at the phase transition, including critical exponent of the correlation length for a trapped, weakly interacting Bose gas \cite{Esslinger}.

In the present publication we focus our attention on a homogeneous Bose gas with attractive interactions in a elongated 3D box with periodic boundary conditions. For a corresponding 1D system the mean field theory (Gross-Pitaevskii equation \cite{GPEcite}) predicts a spontaneous symmetry breaking when the strength of the attractive interaction reaches a critical value \cite{Reinhardt}. On the other hand the one-particle density has to be uniform in the ground state of the full quantum Hamiltonian, due to the translational invariance of the system. The connection between the mean field approach, Bogoliubov \cite{BTcite} and the full quantum many body theory was analyzed by few authors. For instance Ueda and his cooworkers \cite{ueda}, considering 1D Bose gas in the box showed that when the interaction strength exceeds the critical value, even weak external symmetry breaking perturbation leads to a formation of a bright soliton. They also analyzed quantum fluctuations (in position and momentum) of the center of mass of the soliton \cite{Kanamoto}.
Finally, in the context of the present study we mention the paper of Castin and Herzog \cite{Castin} where they suggested that as long as we restrict the analysis of the system to the low order correlation functions it can be characterized quite accurately using modified mean-field approach.

Here we consider quantum many body system and we show that Hilbert space in the vicinity of the critical point can be divided into a subspace, which can be described by the Bogoliubov theory, and a subspace where (two) unstable modes (unstable in the Bogoliubov description) can be very elegantly treated with the help of a so called continuous description. The continuous description in our case leads to a Schr\"odinger equation of a fictitious particle moving in the effective two-dimensional potential and a transformation of this potential from parabolic to Mexican hat shape reflects the second order quantum phase transition. We also use the continuous description to analyze fluctuations close to the critical point in our homogeneous systems. The analogous analysis is also possible for the system in the double well and we will present our results (among other things we show that in this case there is only one unstable mode) in the separate publication \cite{companion}.

The paper is organized as follows. In Sec.~\ref{nbog} we introduce the system and shortly present the results of the Bogoliubov theory. In Section~\ref{contsec} we restrict the analysis to condensate and two unstable modes, using an argument (derived in the previous section) that all  the other modes remain weakly populated. We introduce effective Hamiltonian and continuous description, which allows us to investigate quantum fluctuations in the system in a vicinity of the critical point. We conclude in Sec.~\ref{concl} and in the Appendix we analyze eigenstates of the system and show that for large particle number corrections to an effective Hamiltonian are negligible.

\section{Bogoliubov theory}
\label{nbog}

We consider a gas of Bose particles with contact attractive interaction in a 3D box with periodic boundary conditions. The Hamiltonian of the system reads
\be
\hat H=\int d^3r \, \hat\psi^\dagger(\vect{r})\left[-\frac{\hbar^2}{2m}\nabla^2+\frac{g}{2}
\hat\psi^\dagger(\vect{r})\hat\psi(\vect{r})\right]\hat\psi(\vect{r}),
\ee
where $\hat \psi(\vect{r})$ is the bosonic field operator, and $g=4\pi\hbar^2a/m$, is the coupling constant that characterizes particle interaction ($a$ is $s$-wave scattering length which for the attractive interaction is less than zero). In the case of homogeneous system it is convenient to switch to a momentum basis,
\be
\hat \psi(\vect{r})=\sum_\vect{k} \frac{e^{i\vect{k}\cdot\vect{r}}}
{\sqrt{V}}\hat a_\vect{k},
\ee
where $V=L_xL_yL_z$ is the volume of the box and the sum runs over
discrete momenta,
\be
\vect{k}=2\pi\left(\frac{n_x}{L_x},
\frac{n_y}{L_y},
\frac{n_z}{L_z}\right),
\label{kdis}
\ee
with integer $n_x,n_y,n_z$. Then, the Hamiltonian reads
\be
\hat H=\sum_\vect{k}\frac{\hbar^2k^2}{2m}\hat a_\vect{k}^\dagger\hat a_\vect{k}
+\frac{g}{2V}\sum_{\vect{k},\vect{k'},\vect{q}}
\hat a_{\vect{k}+\vect{k'}-\vect{q}}^\dagger
\hat a_\vect{q}^\dagger\hat a_\vect{k'}\hat a_\vect{k}.
\label{hms}
\ee

Exact diagonalization of the Hamiltonian (\ref{hms}) is impossible because dimension of the Hilbert space of the system is enormous. However, we are interested in a ground state of the system in the case of weak particle interaction. Then, one can expect that only one mode macroscopically occupied by atoms and, in the first approximation, we may use a mean field approach which relies on substitution of the bosonic operator $\hat\psi(\vect{r})$ by a classical field $\phi_0(\vect{r})$. The resulting Gross-Pitaevskii \cite{GPEcite} equation reveals homogeneous stationary solution (i.e. the condensate wavefunction)
\be
\phi_0(\vect{r})=\frac{1}{\sqrt{V}},
\label{hcw}
\ee
and chemical potential $\mu=gN/V$. Quantum fluctuations around the mean field solution can be described within the Bogoliubov theory \cite{BTcite}. We employ $N$-conserving version of the Bogoliubov theory \cite{conserving} where a part of the Hamiltonian (\ref{hms}), minus a constant term $\mu\hat N$, with contributions of the second order in $\hat a_\vect{k}$ (where $\vect{k}\ne 0$),
\bea
\hat H_B&= &
\sum_{\vect{k}\ne 0}\left[\left(\frac{\hbar^2k^2}{2m}+\frac{2g}{V}\hat a_0^\dagger\hat a_0-\frac{gN}{V}\right)
\hat a_\vect{k}^\dagger\hat a_\vect{k}
\right. \cr
&&+ \left.
\frac{g}{2V}\left(\hat a_0^\dagger\hat a_0^\dagger\hat a_{-\K}\hat a_{\K}+
\hat a_0\hat a_0\hat a_{-\K}^\dagger\hat a_{\K}^\dagger
\right)\right],
\label{hamb}
\eea
is substituted by
\bea
\hat H_B&= &
\sum_{\vect{k}\ne 0}\left[\left(\frac{\hbar^2k^2}{2m}+\frac{gN}{V}\right)
\hat \Lambda_\vect{k}^\dagger\hat \Lambda_\vect{k}\right. \cr
&&+ \left.
\frac{gN}{2V}\left(\hat \Lambda_\vect{k}^\dagger\hat \Lambda_{-\vect{k}}^\dagger+
\hat \Lambda_\vect{k}\hat \Lambda_{-\vect{k}}
\right)\right],
\label{hl}
\eea
where
\be
\hat\Lambda_\vect{k}=\frac{\hat a_0^\dagger}{\sqrt{N}}\hat a_\vect{k}.
\ee
The Bogoliubov transformation
$\hat b_\vect{k}=\la u_k|\hat\Lambda_\vect{k}\ra-
\la v_k|\hat\Lambda_{-\vect{k}}^\dagger\ra$, with
\bea
u_k+v_k&=&\left(\frac{\frac{\hbar^2k^2}{2m}}{\frac{\hbar^2k^2}{2m}+\frac{2gN}{V}}\right)^{1/4}, \cr
u_k-v_k&=&\left(u_k+v_k\right)^{-1},
\eea
allows one to write the Hamiltonian (\ref{hl}) in a diagonal form,
$\hat H_B=\sum_{\vect{k}\ne 0}E_k\hat b_\vect{k}^\dagger\hat b_\vect{k}$,
where the Bogoliubov spectrum reads
\be
E_k = \sqrt{\frac{\hbar^2 k^2}{2m}\left( \frac{\hbar^2 k^2}{2m} +  \frac{2gN}{V} \right) }.
\label{bogsp}
\ee
Within the $N$-conserving Bogoliubov theory we can obtain $N$-body ground
state in the particle representation in a simple form \cite{Leggett,dziarmaga}
\be
|0_b\ra\sim\left[\hat a_0^\dagger\hat a_0^\dagger+{\sum_{\vect{k}\ne 0}}
        \lambda_{k}\left(\hat a_{c,\vect{k}}^\dagger\hat a_{c,\vect{k}}^\dagger+
        \hat a_{s,\vect{k}}^\dagger\hat a_{s,\vect{k}}^\dagger\right)\right]^{N/2} |0\ra,
\label{hbg}
\ee
where
\bea
\hat a_{c,\vect{k}}=\frac{\hat a_\vect{k}+\hat a_\vect{-k}}{\sqrt{2}}, \,\,\,\,\,
\hat a_{s,\vect{k}}=\frac{\hat a_\vect{k}-\hat a_\vect{-k}}{i\sqrt{2}},
\label{acs}
\eea
and
\be
\lambda_k=\sqrt{\frac{dN_k}{1+dN_k}}.
\ee
Here
\be
dN_k=\la v_k|v_k\ra,
\ee
are eigenvalues of the reduced single particle density matrix.
The total number of atoms depleted from a condensate mode reads
\be
dN=\sum_{\vect{k}\ne 0}dN_k.
\label{hdn}
\ee
In the following we will consider the system in a box where
$L_x,L_y\le L_z/2$. Then, the lowest Bogoliubov excitation energy corresponds
to momenta
\be
\vect{k_\pm}=\pm\frac{2\pi}{L_z}\left(0,0,1\right),
\label{mpm}
\ee
and it is real
provided a parameter
\be
\alpha=\frac{2gN}{\epsilon_0V},
\label{alpha}
\ee
where
\be
\epsilon_0=\frac{\hbar^2}{2m}\left(\frac{2\pi}{L_z}\right)^2,
\label{e0}
\ee
is greater than $-1$. For $\alpha=-1$ the energy gap between the ground state and the first excited state disappears, the homogeneous mean field solution (\ref{hcw}) looses its stability and the quantum depletion $dN$ diverges because $dN_{k_\pm}$ diverge. More precisely, for $\alpha<-1$ the mean field solution spontaneously breaks the symmetry of the system. In the next section we show that this point can be regarded as a critical point.

We conclude that the Bogoliubov approximation breakdowns down at the critical point. The population of $\K_{\pm}$ is diverging, but population of higher momentum modes is
within this approximation still negligible. Indeed, average number
of particle in these modes for $\alpha$ around $-1$ can be  estimated
as follows:
\be
\sum_{\K\ne 0,\vect{k}_\pm} dN_{k}=\sum_{\K\ne 0,\vect{k}_\pm}\la v_k|v_k\ra \approx
\frac{V}{2\pi^2}\int_{4\pi/L_z}^\infty dk\; k^2
\;\la v_k|v_k\ra,
\label{dnh2}
\ee
and it is not greater than $0.2$ in a vicinity of the critical point for our box geometry.

In the next section we analyze higher order terms of the Hamiltonian (\ref{hms}) and show that one is able to obtain a simple and elegant description of the system in a vicinity of the critical point if the lowest energy modes are carefully considered. Strictly speaking we will consider only three modes, the condensate mode and $\K_{\pm}$ modes.

\section{Continuous Description}
\label{contsec}

In the present section we suggest an effective method to describe our system  (cold gas of bosonic atoms in a box) in a vicinity of the critical point. We consider the elongated rectangular box where $L_x,L_y\le L_z/2$. From the discussion presented above we expect that in this case there are two modes corresponding to momenta $\K_{\pm}$, that need careful treatment.

\subsection{Effective Hamiltonian}

The Hamiltonian restricted to the condensate and $\K_{\pm}$ modes reads
\begin{eqnarray*}
&& \hat H \approx \epsilon_0 \left(\hat a_+^\dagger \hat a_+ + \hat a_-^\dagger \hat a_- \right)+\frac{g}{V} \left(\hat a_0^\dagger \hat a_0^\dagger \hat a_+ \hat a_-  +  \hat a_+^\dagger \hat a_-^\dagger \hat a_{0} \hat a_{0} \right)
\\
\\
&& + \frac{g}{2V} \left( \hat a_0^\dagger \hat a_0^\dagger \hat a_0 \hat a_0 + \hat a_{+}^\dagger \hat a_{+}^\dagger \hat a_+ \hat a_+ + \hat a_-^\dagger \hat a_-^\dagger \hat a_- \hat a_-\right)
\\
\\
&& +\frac{2g}{V} \left(  \hat a_0^\dagger \hat a_0 \left(\hat a_+^\dagger \hat a_+ + \hat a_-^\dagger \hat a_- \right)  + \hat a_+^\dagger \hat a_+ \hat a_-^\dagger \hat a_- \right),
\end{eqnarray*}
where we introduced notation $\hat a_{\K_{\pm}} \equiv \hat a_{\pm}$. Since total number of particles is equal to $N$ and is conserved we can eliminate one of the modes from the above Hamiltonian. Here we choose to eliminate the $k=0$ mode. In order to do so we substitute operators $\hat a_0$ and $\hat a_0^\dagger$ by an operator $\sqrt{  N - \left(\hat a_+^\dagger \hat a_+ + \hat a_-^\dagger \hat a_- \right) }$ (see \cite{Takano}). Now our Hamiltonian can be written as
\begin{equation}
\hat H = \frac{g}{2V}(N^2-N) + \hat H',
\end{equation}
where
\begin{eqnarray*}
&& \hat H' = \left(\epsilon_0 +gn \right)\left( \hat a_+^\dagger \hat a_+ +\hat a_-^\dagger \hat a_-\right) +gn \left( \hat a_+ \hat a_-  +  \hat a_+^\dagger \hat a_-^\dagger \right)
\\
\\
&& -\frac{g}{V} \left(
\hat a_+^\dagger \hat a_+ \hat a_+^\dagger \hat a_+ + \hat a_-^\dagger \hat a_- \hat a_-^\dagger \hat a_- + \hat a_+^\dagger \hat a_+ \hat a_-^\dagger \hat a_- \right)
\\
\\
&&-\frac{g}{V}\left(\left(\hat a_+^\dagger\hat a_+ +\hat a_-^\dagger\hat a_- \right) \hat a_+ \hat a_- + \hat a_+^\dagger \hat a_-^\dagger \left(\hat a_+^\dagger\hat a_+ +\hat a_-^\dagger\hat a_- \right) \right)
\end{eqnarray*}
Note, that the above Hamiltonian commutes with the total momentum operator which in this case is equal to
\begin{equation}
\hat P = \frac{2\pi \hbar}{L_z} \left( \hat a_+^\dagger \hat a_+ - \hat a_-^\dagger \hat a_- \right). \label{Ptotal}
\end{equation}
Next we represent first excited modes by their symmetric and antisymmetric combinations defined in Eq.~(\ref{acs}),
and describe dynamics of these modes using the position-momentum representation
\begin{eqnarray*}
\hat a_c = \frac{\hat x_c + i \hat p_c}{\sqrt{2}}  \ \ \ \ \hat a_s = \frac{\hat x_s + i \hat p_s}{\sqrt{2}}.
\end{eqnarray*}
In this representation the Hamiltonian $\hat H'$ can be split into several parts
\begin{equation}\label{H0}
\frac{\hat H'}{\epsilon_0} = - 1 - \frac{\alpha}{2} - \frac{3\alpha}{8 N} + \frac{\hat H_F}{\epsilon_0}
+ \frac{ \delta \hat H}{\epsilon_0},
\end{equation}
where
\begin{equation*}\label{H1}
\frac{\hat H_{F}}{\epsilon_0}= \frac{1}{2} \left( \hat p_c^2 + \hat p_s^2 \right)   + \frac{ 1 + \alpha }{2}\left(\hat x_c^2 + \hat x_s^2\right)-\frac{7\alpha}{32 N} \left( \hat x_c^2 + \hat x_s^2 \right)^2
\end{equation*}
and
\begin{eqnarray*}\label{dH1}
&& \frac{\delta \hat H}{\epsilon_0} =  -\frac{\alpha}{8N} \left(\hat p_c^2 + \hat p_s^2\right)   + \frac{7\alpha}{8N}  \left(\hat x_c^2 + \hat x_s^2\right)
\\
\\
&& -\frac{3\alpha}{32 N}\left( ( \hat p_c^2 + \hat p_s^2)(\hat x_c^2 + \hat x_s^2) +  (\hat x_c^2 + \hat x_s^2)( \hat p_c^2 + \hat p_s^2) \right)
\\
\\
&& -\frac{\alpha}{32 N} \left(4\left( \hat p_c \hat x_s - \hat p_s \hat x_c  \right)^2  - \left( \hat p_c^2 + \hat p_s^2 \right)^2 \right).
\end{eqnarray*}
The total momentum operator (\ref{Ptotal}) is in this representation proportional to $\hat p_c \hat x_s - \hat p_s \hat x_c$ and it commutes with the Hamiltonian $\hat H_F$ (and obviously also with $\delta \hat H$)! This property will be used in the Appendix.

Up to now we have divided our Hamiltonian into two parts and in what follows we will argue that for large $N$ (strictly speaking in the limit of $N$ tending to infinity while $\alpha$ is kept constant) the dominating contribution arise from effective Hamiltonian $\hat H_F$. This is the Hamiltonian of a fictitious particle moving in a two-dimensional
effective potential. Indeed, upon defining $\hat r^2 = \hat x_c^2 + \hat x_s^2$  we can derive a Schr\"odinger equation
\be
E \psi(r)= - \frac12  \triangle \psi(r) + V_{eff}(r)\psi(r),
\label{scheff}
\ee
where the 2D effective potential is equal to
\be
V_{eff}(r)=\frac{1 +\alpha}{2} r^2- \frac{7\alpha}{32N} r^4.
\label{veff}
\ee
This potential evolves, when we pass through the critical point, from a parabolic well ($\alpha>-1$), through quartic well ($\alpha=-1$) to a Mexican hat shape ($\alpha<-1$), which is a signature of the second order quantum phase transition. In the Appendix we present analytic approximations to the solution of this Schr\"odinger equation and estimate the contribution from the Hamiltonian $\delta \hat H$ in all three regions mentioned above. We show that for large $N$ they are indeed negligible.

\begin{figure}[h]
\includegraphics[scale=0.7,angle=270,clip]{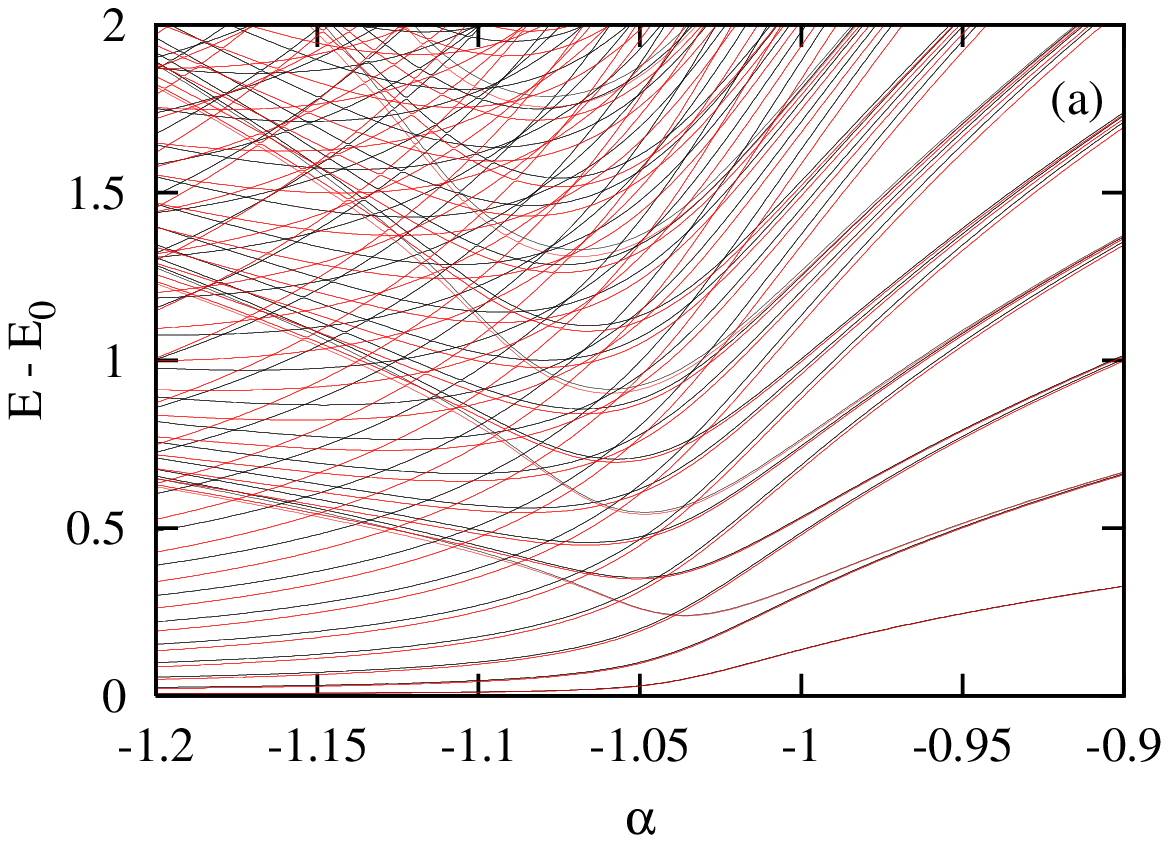}
\includegraphics[scale=0.7,angle=270,clip]{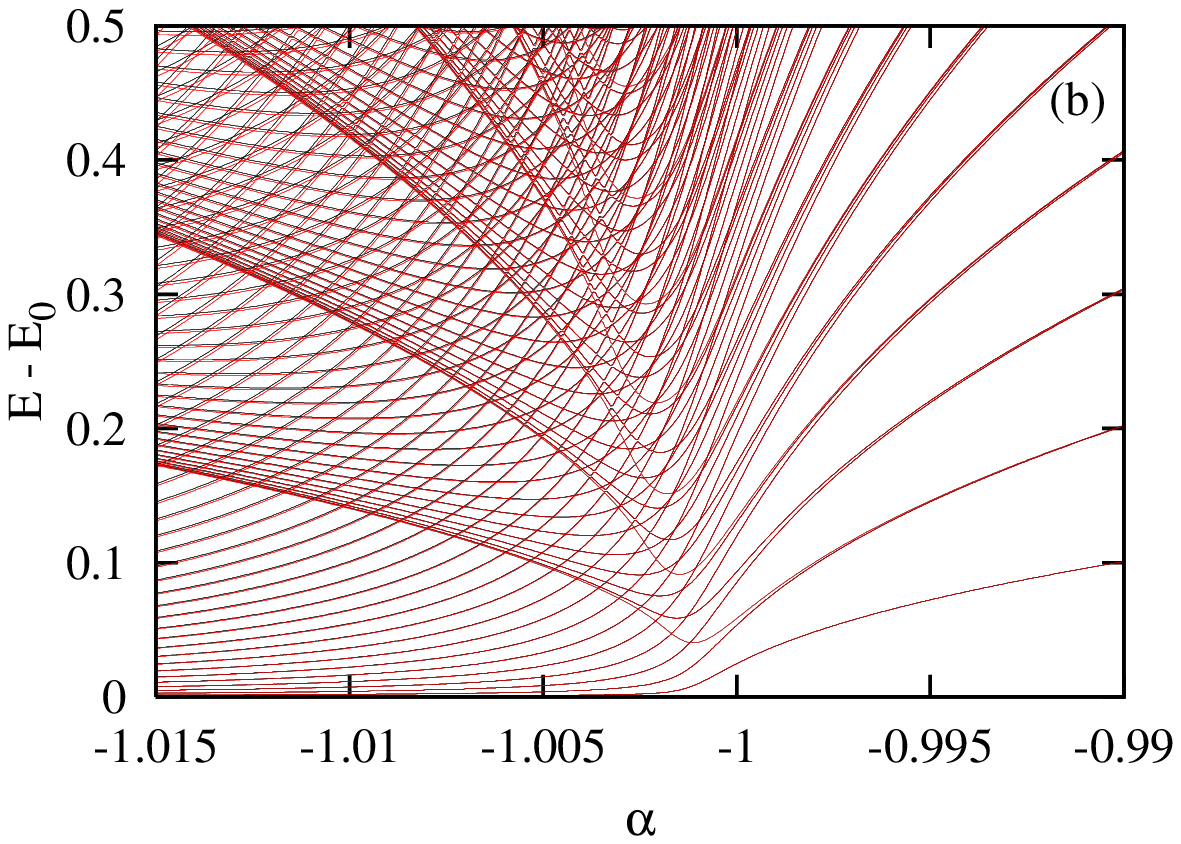}
\caption{(Color online) Energy levels with respect to the ground state energy of the Hamiltonian $H'$ (black lines) and $H_F$ (red lines). Panel (a) shows results for $N=500$ while panel (b) for $N=10^5$.}
\label{fig1}
\end{figure}

Our numerical studies are summarized in Fig.~\ref{fig1}. It turns out that our estimates from the Appendix are in a perfect agreement with the numerical results. To prove that the contribution from $\delta \hat H$ is small we diagonalize Hamiltonian $\hat H'$ and $\hat H_F$ numerically. In Fig.~\ref{fig1} we show energy levels obtained in each case. One can see that already for a moderate particle number ($N=500$) the low-lying energy levels of the two sets coincide and for the greater $N$ ($N=10^5$) even the higher excited states are properly reproduced. In conclusion even for moderate number of particles Hamiltonian $\hat H_F$ is a very good approximation to $\hat H'$ apart from the constant term $ - 1 - \frac{\alpha}{2} - \frac{3\alpha}{8 N}$.

\subsection{Critical fluctuations}

Description of a homogeneous Bose gas with attractive particle interactions around the critical point becomes very simple when we use the concept of the effective potential (\ref{veff}). For instance it allows us to study the dynamical transition of the system when the scattering length is changed in time, especially when the system enters the critical region. Here, however, we restrict our considerations to the static (ground state)
properties of the system around the critical point. Since the effective potential undergoes transition from quadratic through quartic to Mexican hat, which is a typical signature of the second order phase transition, we expect critical fluctuations to show up. Hence we search for the observable that will show the clear evidence of the critical behavior (will have maximal fluctuations at the critical point).

\begin{figure}[h]
\includegraphics[scale=0.4]{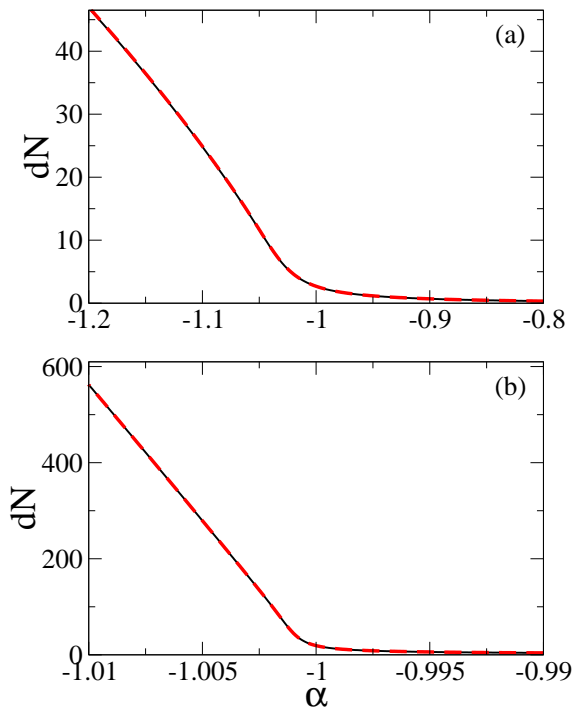}
\caption{(Color online) Mean depletion from the condensate as a function of $\alpha$ for $N=500$  (a) and $N=10^5$  (b). Black continuous lines correspond to the Hamiltonian $\hat H'$ and red dashed lines to $\hat H_F$.}
\label{fig3}
\end{figure}

Let us consider the operator $d\hat N$ representing the number of atoms depleted from a condensate. The contribution (\ref{dnh2}) from the modes described by the Bogoliubov Hamiltonian is very small and can be neglected, therefore
\be
d\hat N\approx \hat n_c + \hat n_s,
\ee
where
\be
\hat n_c = \frac{1}{2}\left( \hat p_c^2 +\hat x_c^2 -1 \right),
\quad
\hat n_s = \frac{1}{2}\left( \hat p_s^2 +\hat x_s^2 -1 \right). \label{sc}
\ee
The depletion, $dN=\la d\hat N\ra$, increases as we approach and pass through the critical point, see Fig.~\ref{fig3}. In this figure we compare mean value of atoms depleted from the condensate as a function of parameter $\alpha$ for two different total number of atoms. Each panel represents two curves that practically sit on top of each other; one obtained from
the Hamiltonian $\hat H'$ and the other from $\hat H_F$. The linear behavior is due to the linear dependence of the square of the position of the minimum in the Mexican hat --- $r_0^2$ on the interaction strength $\alpha$, (see the Appendix). For large $N$ we can estimate that
$\langle \hat n_c + \hat n_s \rangle \simeq  \langle \hat x_c^2 + \hat x_s^2 \rangle /2 \simeq r_0^2/2 = N\frac{4(\alpha+1)}{7\alpha}$.

\begin{figure}[h]
\includegraphics[scale=0.4]{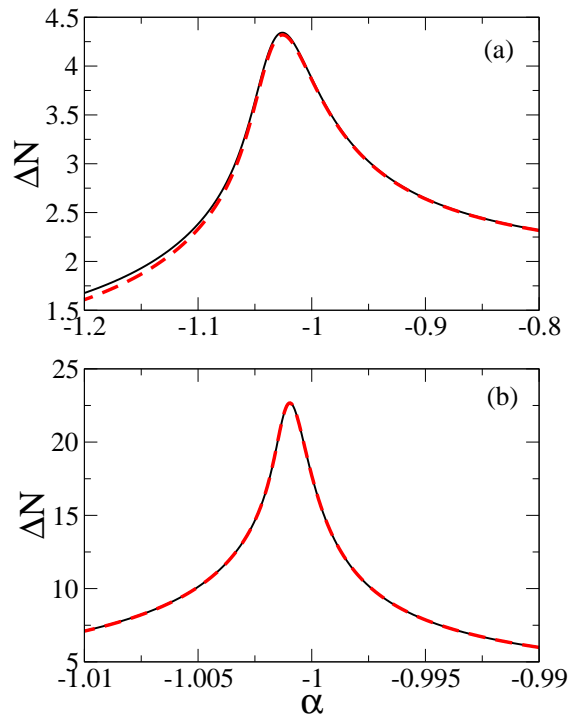}
\caption{(Color online) Fluctuations of the condensate depletion (\ref{fl}) for $N=500$  (a) and $N=10^5$  (b) as a function of $\alpha$. Black continuous lines correspond to the Hamiltonian $\hat H'$ and red dashed lines to $\hat H_F$. }
\label{fig4}
\end{figure}

Now we carry on to the fluctuations. The variance of the $d\hat N$ increases as we cross the critical point but if we calculate the variance relative to the average value it turns out that the critical point region is clearly indicated by the maximum of such fluctuations.
In Fig.~\ref{fig4} we present the variable
\be
\Delta N=\frac{\langle (d \hat N - \langle d \hat N \rangle)^2 \rangle}{\langle d \hat N \rangle},
\label{fl}
\ee
as a function of $\alpha$ around the critical point. We compare again two curves obtained using Hamiltonians $\hat H'$ and $\hat H_F$ for two different number of particles. The agreement is already satisfactory for 500 particles and it is excellent in the case of $10^5$ particles. We observe that the maximum is shifted from the point $\alpha = -1$ and this shift tends to zero with the increasing number of particles.

It is interesting to note  that for $\alpha<-1$ the fluctuations of $d\hat N$ are much smaller than the fluctuations of $\hat n_c$ and $\hat n_s$ separately. Indeed, for $\alpha<-1$, the ground state solution corresponds to the wavefunction concentrated around a circle with a radius $r_0$ [see Eq.~(\ref{r0})], and because $d\hat N$ is a function of the distance from the origin its fluctuations are small. The operators $\hat n_c$ and $\hat n_s$ are related to distances from the axes, and their fluctuations are larger than the fluctuations of $d\hat N$ if the wavefunction is localized around the ring. It means that while the number of depleted atoms is expected to be roughly the same in each experiment, these atoms may occupy differently two orthogonal modes, i.e. the modes (\ref{mpm}) or any other orthogonal combination of them. This is the origin of spontaneous symmetry breaking and bright soliton formation, i.e. atoms depleted from a condensate in different experiments may differently occupy the modes (\ref{mpm}) and the resulting density profiles of atomic clouds reveal bright solitons localized at different positions.

\section{Conclusions}
\label{concl}

We consider a ground state of the Bose gas with attractive particle interactions in an elongated 3D box. When strength of these interactions increases the system passes through a critical point, which is indicated by loss of stability of the uniform condensate wavefunction and breakdown of the Bogoliubov approach. In the vicinity of the critical point there are {\it two unstable modes} and {\it they} require a careful
treatment --- all other modes can be neglected using argument, emerging from
Bogoliubov theory, that they do not get any significant population.

We introduce an effective hamiltonian and the continuous description which allows for a simple analysis of the unstable modes subsystem. It turns out that the subsystem can be effectively described by a 2D Schr\"odinger equation of a fictitious particle moving in an effective potential. This potential changes when the interaction strength increases and it evolves from a parabolic, through quartic (at the critical point) to a Mexican hat shape which reflects the second order quantum phase transition.

We have analyzed properties of the ground state, energy levels structure and quantum fluctuations of the system in a vicinity of the critical point. The fluctuations of number of atoms depleted from a condensate (relative to an average value) turn out to be maximal at the critical point. These fluctuations are smaller than fluctuations of numbers of atoms occupying each of the unstable modes separately. It means that the depleted atoms may very differently occupy two unstable modes in different experimental realizations. This is an origin of the spontaneous symmetry breaking and bright soliton formation that is expected to be observed in a single experimental measurement.

\section*{ Acknowledgements }
The work of BO and KS was supported by Polish Government scientific funds
(2005-2008) as a research project. Supported by Marie Curie ToK project
COCOS (MTKD-CT-2004-517186). P. Z. and M. T. acknowledge support by Polish Government scientific funds (2007-2010).


\section{APPENDIX: Eigenstates of the effective Hamiltonian}

Here we present arguments based on the analytic analysis to demonstrate that Hamiltonian $\delta \hat H$ can be neglected in comparison with $\hat H_F$ for large $N$ when we consider the ground state of the system. The procedure is as follows. First we find a ground state of $\hat H_{F}$ (we treat it separately in the three regions depending on the value of $\alpha$). With some straightforward approximations we are able to find analytical expressions for the ground state wavefunction and estimate the energy difference $\Delta E$ between the ground and the first excited state (with total momentum equal to zero, see below).  Next we estimate the contribution of the Hamiltonian $\delta \hat H$ using perturbation theory and show that for large $N$ first order correction is much smaller than $\Delta E$.

Notice that as we mentioned above total momentum operator (\ref{Ptotal})commutes with both $\hat H_{F}$ and $\delta \hat H$. Hence any excitation caused by the latter can not change the value of the total momentum, which is equal to zero in the ground state. If we use continuous description and express the total momentum operator in the polar coordinates, it turns out that it is proportional to $\partial /\partial \phi$. Hence any state with total momentum equal to zero do not depend on $\phi$. In conclusion, to estimate $\Delta E$ we neglect the $\phi$ dependence.

Finally we stress that considerations presented here are given only as an example of the back of the envelope approximation, since in parallel we perform numerical calculations and obtain exact energies and eigenstates of both Hamiltonians.

\bi
\item
{\it The $\alpha +1>0$ case.}

In this case in order to get an analytic expression for the ground state we neglect the fourth order terms in the effective potential (\ref{veff}). Then the Schr\"odinger equation (\ref{scheff}) can be written in the form
\be
 - \frac12\triangle \psi(r)
+ \frac{1 + \alpha}{2} r^2 \psi(r) = E \psi(r),
\label{sch1}
\ee
and the ground state solution is
\begin{equation}
\psi (r) \sim \exp \left( -\frac{r^2}{2 a_{ho}^2} \right),
\end{equation}
where
\begin{equation}
a_{ho}^4 = \frac{1}{1+\alpha}.
\end{equation}
The difference $\Delta E$ between the energy levels is of the order of $a_{ho}^{-2}$. The fourth order terms omitted in Eq.~(\ref{sch1}) result in corrections of the order of $a_{ho}^4/N$.
One can  omit the fourth order terms in the effective potential when $ a_{ho}^4/N \ll a_{ho}^{-2}$.
Inserting the value of $a_{ho}$ from the above equation we get \cite{remark}
\begin{equation}\label{cond1}
 (1+\alpha)^{3/2}  \gg \frac{1}{N}.
\end{equation}
The biggest term in the Hamiltonian $\delta \hat H$ is of the order $a_{ho}^2/N$ and is  smaller than
just discussed fourth order term. So in the region where the above solution is valid the Hamiltonian $\delta \hat H$ has negligible impact.

\item
{\it The critical point, $\alpha =-1$.}

For $\alpha=-1$ the Schr\"odinger equation (\ref{scheff}) takes the form
\be
 - \frac12 \triangle \psi(r) +\frac{7}{32 N} r^4 \psi(r) = E \psi(r).
\ee
We switch to the new variable $ \tilde r = r/N^{1/6}$ and obtain
\be
 -\frac12 \triangle \psi(\tilde r)
+ \frac{7}{32}  \tilde r^4 \psi(\tilde r) = N^{1/3}E \psi( \tilde r).
\ee
From dimensional analysis we estimate $\Delta E$ to be of the order of $N^{-1/3}$.
On the other hand the dominant contribution to the Hamiltonian $\delta \hat H$  comes from the terms $\hat x^2/N$ and is of order of $N^{-2/3}$. So in the limit of large $N$ it can be neglected.


\item
{\it The $\alpha+1 < 0$ case.}

For $\alpha<-1$ the potential in Eq.~(\ref{veff}) has the shape of a Mexican hat. The minimum occurs for
\begin{equation}
r_0^2 = N\frac{8(\alpha+1)}{7\alpha}.
\label{r0}
\end{equation}
Around this minimum, we choose the variable $R = r - r_0$, and rewrite the Schr\"odinger equation (\ref{scheff}) as
\begin{eqnarray}\label{p2}
E \psi(R) &=& - \frac{1}{2}\left(\frac{\mbox{d}^2}{\mbox{d} R^2}
+ \frac{1}{r_0+R}\frac{\mbox{d}}{\mbox{d} R} \right) \psi(R)
\\ \nonumber
\\ \nonumber
&&-(1+\alpha) R^2 \psi(R) - \frac{7 \alpha}{32 N} R^4 \psi(R),
\end{eqnarray}
use harmonic approximation (see the comment below) and finally obtain
\begin{eqnarray}
 - \frac{1}{2}\frac{\mbox{d}^2}{\mbox{d} R^2}
 \psi(R)
-(1+\alpha) R^2 \psi(R)  = E \psi(R). \label{harmon}
\end{eqnarray}
The ground state is
\begin{equation}
\psi(R) \sim \exp \left( - \frac{R^2}{2b_{ho}^2} \right),
\end{equation}
where
\begin{equation}
b_{ho}^4 = -\frac{1}{2(1+\alpha)}.
\end{equation}
Hence, the energy level spacing $\Delta E$ is proportional to $b_{ho}^{-2}$.
To justify our approximation leading to the harmonic equation  (\ref{harmon}) we notice that the omitted terms in equation (\ref{p2}) are of the order $r_0^{-1} b_{ho}^{-1}$ and $b_{ho}^4/N$.
These terms are negligible when
\begin{equation}
|1+\alpha|^{3/2} \gg \frac{1}{N}.
\end{equation}
Incidentally it is the same condition as (\ref{cond1}) obtained earlier.

Even if the harmonic approximation of equation (\ref{p2}) is valid we still need to estimate the
contribution of $\delta \hat H$. First order perturbation gives
$b_{ho}^{-2} r_0^2/N$ as a leading order.
Hence, comparing it with the distance between the energy levels, we obtain the condition when the Hamiltonian
$\delta \hat H$ can be neglected to be
\begin{equation}
|1+\alpha |  \ll 1
\end{equation}
\ei



\begin{thebibliography}{99}
\bibitem{remark} This condition sets the range of $\alpha$ where the entire system may be
described by the Bogoliubov theory.
\end{thebibliography}
\end{document}